\documentclass{article}
\usepackage{epsfig}
\usepackage{amssymb}
\usepackage{amsmath}
\usepackage{amsfonts}
\newcommand \be{\begin{eqnarray}}
\newcommand \ee{\end{eqnarray}}
\begin{document}
\begin{center}
{\bf Real-Time Kadanoff-Baym Approach to Nuclear Response
Functions}\\

\bigskip
\bigskip
H. S. K\"ohler \footnote{e-mail: kohler@physics.arizona.edu} \\
{\em Physics Department, University of Arizona, Tucson, Arizona
85721,USA}\\
N. H. Kwong \footnote{e-mail:kwong@optics.arizona.edu} \\
{\em College of Optical Sciences, University of Arizona,
Tucson, Arizona 85721, USA}  \\

\end{center}
\date{\today}

\begin{abstract}
Linear density response functions are calculated for symmetric nuclear matter
of normal density by time-evolving two-time Green's functions in real
time. Of particular interest is the effect of correlations.
The system is therefore initially time-evolved with a collision term
calculated in a direct Born approximation
as well as with full (RPA) ring-summation until fully correlated. An
external time-dependent potential is then applied. The ensuing density
fluctuations are recorded to calculate the density response.
This method was previously used by Kwong and Bonitz for studying plasma
oscillations in a correlated electron gas.
The energy-weighted sum-rule for the response function is guaranteed by
using
conserving self-energy insertions as the method  then generates
the full vertex-functions. These can alternatively be calculated by
solving a Bethe -Salpeter equation as done in some previous works.
The (first order) mean field is derived
from a momentum-dependent (non-local) interaction while
$2^{nd}$  order self-energies  are calculated  using  a particle-hole
two-body effective (or 'residual')  interaction given by a
gaussian \it local \rm  potential.

We present numerical results for the response function $S(\omega,q_0)$ for
$q_0=0.2,0.4$ and $0.8 {\rm fm}^{-1}$. Comparison is made with the nucleons being
un-correlated i.e. with only the first order mean field included, the 'HF+RPA'
approximation.

We briefly discuss the relation of our work with the Landau quasi-particle
theory as applied to nuclear systems by Sj\"oberg and followers using  methods
developped by Babu and Brown, with special emphasis on the 'induced' interaction.

\end{abstract}


\vskip2pc

\section{Introduction}
Response functions, the response of a many-body system to an external
perturbation is instrumental in our understanding of the properties and
interactions involved in the excitations of the system.
In the study of nuclear systems these response functions are of particular
interest when it comes to calculate the mean free path and absorpton of e.g.
neutrinos in a neutron gas \cite{mar04,iwa82}, a subject of interest in astrophysical
studies.\cite{red98,sed00}
%

This report concerns the calculation of  linear density response functions
for symmetric nuclear matter. 
The excitations of the medium in response to an external perturbation is of course
closely dependent on the interactions in the medium.
Of particular interest here is the effect of correlations
 i.e. dressed propagators and collision-terms.
It was shown by  Baym and Kadanoff \cite{bay62}
that, if one wishes to construct the linear response
function with dressed equilibrium Green's functions,
appropriate vertex corrections to the polarization bubble
are necessary to guarantee the preservation of the local
continuity equation for the particle density and current
in the excited system,
which in turn implies the satisfaction of the energy-weighted
sum-rule.
This problem has traditionally involved solving  a Bethe-Salpeter
equation to calculate  these vertex corrections.
Such a calculation was done by Bozek for nuclear
matter\cite{boz04} and further applied in ref.\cite{boz05} while
Lykasov et al used the Landau transport equation with a collision term and vertex
corrections.\cite{lyk04}
Response functions have of course since long been the focus of intense
studies related to the electron gas. Early works
by Lundqvist and Hedin are noticeable leading to the often cited GW
method. \cite{hed65,hed69}
More recent works include those of Faleev and Stockman focussing on
electrons in quantum wells.\cite{fal01}

An alternative method, that does not require an explicit solution of the 
Bethe-Salpeter equations,
is a real time solution of the Kadanoff Baym two-time equations equations using
conserving approximations for the  self-energies. This method
guarantees that  the energy weighted sum-rule (assuming an effective mass $m^*$)
$\int_{0}^{\infty}\omega S(\omega,q_0)d\omega =\frac{q_0^2}{2m^*}$ is satisfied.
It was first used by Kwong and Bonitz for the plasma oscillations in an electron 
gas \cite{nhk00} and later in ref.\cite{bon01}.
They presented a fully linearised formalism separating the Green's function  
into a spatially homogeneous part $G_{00}$ and a linear response part $G_{10}$.
As made clear in ref. \cite{bon99} there is no need to separately
calculate vertex corrections when applying this formalism.  The proper vertex 
corrections  are generated by the time-evolution of the Green's functions when 
subjected to the external disturbance.
We use this  method  here  to calculate density  response functions in symmetric 
nuclear matter.

The relevant KB-equations, already presented in ref. \cite{nhk00} are shown in 
section 2. followed in sect 2.1 by an expos\'e and explanation of 
the interactions used at the separate level of approximations.  
When  the system initially correlates with the total 
energy being conserved, the potential energy of interactions decreases and the
kinetic energy increases, resulting in an increase of  the temperature.
Section 2.2 shows how  to calculate the resulting temperature after equilibration.
Our main computed results are shown in Sect. 3.  The relation of of our work to the Landau
quasi-particle theory is addressed in Sect. 4 and Sect. 5 contains a summary 
of our results and some conclusions.
A diagrammatic representation of the main equations is shown in an Appendix.

\section{The two-time KB-equations}
The Kadanoff-Baym equations in the two-time form for the specific
problem at hand was already shown in previous work, where it was applied
to the electron gas in a linear response calculation.\cite{nhk00}
Some modifications (related to the difference in  interactions and self-energies) 
are necessary for the present nuclear problem.
Correlations are included by  defining self-energies  in a second order Born
approximation with  a residual interaction defined by eq.(\ref{eq3bb}) below.
This is in addition to a first order mean (Hartree-Fock) field.
This then is a 'conserving approximation'  \cite{bay62} a necessary
requirement when applying the two-time formalism consistently.
We compare our results with calculations where correlations are neglected with only
the mean field included, the 'HF+RPA' approximation.

We consider three separate cases: \\
I.  Uncorrelated, mean field only, RPA approximation.\\
II. Correlations included by  self-energies  in second order Born of  a
residual interaction, eq.(\ref{eq3bb}) below. \\
III. Correlations modified by  including not only second order diagrams but 
ring-diagrams  to all orders (i.e.  RPA) in the self-energies. \\
The selfenergy obtained with the second order Born approximation as well as RPA
are 'conserving approximations'  \cite{bay62} a necessary
requirement when applying the two-time formalism consistently.

Calculations proceed as follows:
Equilibrium Green's functions are first constructed for an uncorrelated fermi
distribution of specified density and temperature. These functions, $G_{00}$,
are then time-evolved (for typically $10 fm/c$) with the chosen mean field and
correlations (I,II or III above) until stationary.  
An external field $U({\bf q},t)=U_{0}(t)\delta_{{\bf q,q_{0}}}$ is then applied
which  generates particle-hole Green's functions $G^{^{>}_{<}}_{10}$ that propagate 
in time according to equations  already shown in ref. \cite{nhk00} and for 
completeness repeated here (summation over $m=0,1$ and integration over 
$\bar{t}$ from $-\infty$ to $+\infty$ is implied).

\begin{eqnarray}
\left(i\hbar \frac{\partial}{\partial t}-\epsilon_{{\bf k+q_0}} \right )
G_{10}^{^{>}_{<}}({\bf k}tt')=U_0(t)G_{00}^{^{>}_{<}}({\bf k}tt')
+\Sigma_{1m}^{HF}({\bf k}t)G_{m0}^{^{>}_{<}}({\bf k}tt')
\nonumber \\
+\Sigma_{1m}^{R}({\bf k}t\bar{t})G_{m0}^{^{>}_{<}}({\bf k}\bar{t}t')+
\Sigma_{1m}^{^{>}_{<}}({\bf k}t\bar{t})G_{m0}^{A}({\bf k}\bar{t}t')
\label{eq01a}
\end{eqnarray}
and
\begin{eqnarray}
\left(-i\hbar \frac{\partial}{\partial t'}-\epsilon_{{\bf k}} \right )
G_{10}^{^{>}_{<}}({\bf k}tt')=U_0(t')G_{11}^{^{>}_{<}}({\bf k}tt')
+G_{1m}^{^{>}_{<}}({\bf k}tt')\Sigma_{m0}^{HF}({\bf k}t')
\nonumber \\
+G_{1m}^{R}({\bf k}t\bar{t'})\Sigma_{m0}^{^{>}_{<}}({\bf k}\bar{t}t')
+G_{1m}^{^{>}_{<}}({\bf k}t\bar{t'})\Sigma_{m0}^{A}({\bf k}\bar{t}t')
\label{eq01b}
\end{eqnarray}

We point out that $G^{^{>}_{<}}_{10}$ 'carries' two momenta; 
$G^{^{>}_{<}}_{10}({\bf k}tt')\equiv G^{^{>}_{<}}_{10}({\bf k+q_0},t;{\bf k},t')$.
\cite{nhk00}

In case III
the rings are included in the selfenergies $\Sigma_{00}^{^{>}_{<}}$  by
\begin{eqnarray}
\Sigma_{00}^{^{>}_{<}}({\bf k},t,t')=
i\sum_{\bf p}
G_{00}^{^{>}_{<}}({\bf k-p},t,t')
V_{s}^{^{>}_{<}}({\bf p},t,t').
\label{eqring}
\end{eqnarray}
with
\begin{eqnarray}
V_{s}^{^{>}_{<}}({\bf p},t,t') =
V({\bf p})[\int_{t_{0}}^{t} dt''(\Pi_{00}^{>}({\bf p},t,t'')-
\Pi_{00}^{<}({\bf p},t,t''))V_{s}^{^{>}_{<}}({\bf p},t'',t')-
\nonumber \\
\int_{t_{0}}^{t'} dt''\Pi^{^{>}_{<}}({\bf p},t,t'')
(V_{s}^{>}({\bf p},t'',t')-V_{s}^{<}({\bf p},t'',t'))]+ \nonumber\\
V^{2}({\bf p})\Pi_{00}^{^{>}_{<}}({\bf p},t,t')
\label{eq1}
\end{eqnarray}
where $V({\bf p})$ is the momentum-representation of the residual potential,
local in ccordinate space, eq. (\ref{eq3bb}).
The polarisation bubble $\Pi_{00}$ is defined by
\begin{eqnarray}
\Pi_{00}^{^{>}_{<}}({\bf p},t,t')=
-i\sum_{\bf p'}
G_{00}^{^{>}_{<}}({\bf p'},t,t')
G_{00}^{^{<}_{>}}({\bf p'-p},t',t)
\label{eq3c}
\end{eqnarray}
In a second order calculation (case II) only the last term (one bubble) in
eq. (\ref{eq1}) will contribute to $V_{s}$.

The polarisation bubble in the $(10)$-channel is given by
\begin{eqnarray}
\Pi_{10}^{^{>}_{<}}({\bf p},t,t')=
-i\sum_{\bf p'}
[G_{10}^{^{>}_{<}}({\bf p'},t,t')
G_{00}^{^{<}_{>}}({\bf p'-p},t',t) \nonumber \\
+G_{00}^{^{>}_{<}}({\bf p'},t,t')
G_{10}^{^{<}_{>}}({\bf p'-p-q_0},t',t)]
\label{eq3d}
\end{eqnarray}

and the selfenergies in the ({10}) channel are given by
\begin{eqnarray}
\Sigma_{10}^{^{>}_{<}}({\bf k},t,t')=
i\sum_{\bf p}
[G_{10}^{^{>}_{<}}({\bf k-p},t,t')
V_{s}^{^{>}_{<}}({\bf p},t,t')     \nonumber \\
+G_{00}^{^{>}_{<}}({\bf k-p},t,t')
V_{s(10)}^{^{>}_{<}}({\bf p},t,t')]
\label{eq4}
\end{eqnarray}

In case III where rings are included to all orders one has

\begin{eqnarray}
V_{s(10)}^{^{>}_{<}}({\bf p},t,t')= a_{10}^2
\int_{t_{0}}^{t}d\bar{t}
\int_{t_{0}}^{t'}d\bar{t'} \nonumber   \\
\{ V_s^{R}({\bf p+q_0},t,\bar{t})
\Pi_{10}^{R}({\bf p},\bar{t},\bar{t'})
V_s^{^{>}_{<}}({\bf p},\bar{t'},t') \nonumber  \\
+V_s^{R}({\bf p+q_0},t,\bar{t})
\Pi_{10}^{^{>}_{<}}({\bf p},\bar{t},\bar{t'})
V_s^{A}({\bf p},\bar{t'},t')  \nonumber \\
+V_s^{^{>}_{<}}({\bf p+q_0},t,\bar{t})
\Pi_{10}^{A}({\bf p},\bar{t},\bar{t'})
V_s^{A}({\bf p},\bar{t'},t') \}
\label{eq4a}
\end{eqnarray}
See following section regarding the $a_{10}$-factor and interactions.
The retarded and advanced parts are given by

\begin{eqnarray}
\Sigma_{10}^{R/A}({\bf p},t,t')=\pm \theta(\pm (t-t')[\Sigma_{10}^{>}({\bf
p},t,t')-\Sigma_{10}^{<}({\bf p},t,t')]
\label{eq5a}
\end{eqnarray}
\begin{eqnarray}
\Pi_{10}^{R/A}({\bf
p},t,t')=\pm\theta(\pm (t-t')[\Pi_{10}^{>}({\bf
p},t,t')-\Pi_{10}^{<}({\bf p},t,t')]
\label{eq5b}
\end{eqnarray}
\begin{eqnarray}
V_{s}^{R/A}({\bf p},t,t')=\delta(t-t')V({\bf p})
\pm\theta(\pm (t-t')[V_{s}^{>}({\bf
p},t,t')-V_{s}^{<}({\bf p},t,t')]
\label{eq6b}
\end{eqnarray}

In case II $V_{s(10)}$ simplifies to
\begin{equation}
V_{s(10)}^{^{>}_{<}}({\bf p},t,t')=
V_a({\bf p})V_a({\bf p+q_0})\Pi_{10}^{^{>}_{<}}({\bf p},t,t')
\label{eq7b}
\end{equation}

with $V_a$ specified below in section 2.1.

For the Hartree-Fock self-energy in the ($00$)-channel 
we choose a parametrisation shown in section 2.1 below.
The Hartree-Fock self-energy in  the ($10$)-channel  is given by

\begin{equation}
\Sigma_{10}^{HF}({\bf p},t)=-iV_a(q_0)\sum_{\bf p'}
 G^{<}_{10}({\bf p'},t,t)+i\sum_{\bf p'}G^{<}_{10}({\bf
 p-p'},t,t)V_a(p')
\label{HF01}
\end{equation}

A diagrammatic representation of the above
equations is shown in the Appendix.

The eqs. (\ref{eq01a}) and (\ref{eq01b}) are time-evolved and the
response-function is, after a fourier-transformation of
$$ \delta n(q_0,t)=-i\sum_{\bf p}G^{<}_{10}({\bf p},t,t)$$
with respect to time t, calculated from
\begin{equation}
S(\omega,q_0)=\frac{\delta n(q_0,\omega)}{\pi n_{0}U_{0}(\omega)}\delta n(q_0,\omega)
\label{S}
\end{equation}

\subsection{Interactions}
The nuclear two-body problem in 'free space' is still unresolved in
detail  but it involves a combination of long ranged   one-pion exchange
contributions and complicated strong short ranged interactions and repulsions.
This leads to even more complicated 'in-medium' interactions and
strong correlations  in the nuclear many-body environment. Simplifications
are necessary to make the exploratory calculations  presented here
reasonably short.
The nuclear  model adopted  in most similar works is that of nucleons
moving in a mean (' Hartree-Fock self-consistent') field with mutual
in-medium ('residual') two-body interactions defined by some potential.
A reasonable form of this potential could be
a Skyrme- ( as used in ref. \cite{gar92}) or the Gogny-interaction
( as used in ref.  \cite{gog77,mar05})  or some  similar interaction.
Apart from these in-medium effective potentials, a
semi-realistic NN-interaction with a hard core was used in a
Jastrow-correlated extension of the RPA.\cite{kwo84}.
We are taking a  somewhat different approach here
adopting three different potentials (A,B and C)`
 each representing  different aspects (domains) of the interaction
that for the purpose of presentation  we assume to be  a Brueckner
state-independent reaction matrix $<{\bf k}|K|{\bf k'}>$ or
$K({\bf p,q})$ with $q={\bf k-k'}$ and $p=\frac{1}{2}({\bf k+k'})$.

A. An important feature of the Hartee-Fock mean field is that it is
momentum-dependent and that only the diagonal matrix-elements $K(p,0)$ are needed
for its calculation.
We  adopt a parametrisation due to  Welke et al \cite{wel88}
leading to:

\begin{equation}
\Sigma^{HF}_{00}({\bf p},t)=A\frac{\rho}{\rho_o}+
B\left(\frac{\rho}{\rho_0}\right)^{\sigma}+
2\frac{C}{\rho_o}
\int \frac{\rho({\bf p'},t)}
{1+(\frac{{\bf p-p'}}{\Lambda})^2}
\frac{d {\bf p'}}{(2\pi)^3}
\label{VG}
\end{equation}
with $A=-110.44$ MeV, $B=140.9$ MeV, C=-64.95 MeV, $\sigma=1.24$,
$\rho_o=0.16 {\rm fm}^{-3}$ , $\Lambda=1.58p_F$
while $\rho$ is the density of the symmetric nuclear matter under
consideration and $\rho({\bf p},t)$ is defined below.

We shall also, in separate calculations substitute this mean field by
assuming an effective mass.

B. To calculate the second order Born and ring diagrams we need
off-diagonal matrix-elements of the interaction i.e. its $q$-dependence.
We choose here an interaction often used in this context and first introduced by
Danielewicz \cite{boz04,boz05,dan84,hsk95}
given by:
\begin{equation}
V(r)=V_0  e^{-(\frac{r}{\eta})^2}
\label{ggg}
\end{equation}
with $\eta=0.57 {\rm fm}$ and $V_0=-453$ MeV.

In momentum-space it reads:
\begin{equation}
V({\bf q})=\pi^{3/2}\eta^{3}V_{0}e^{-\frac{1}{4}\eta^{2}q^{2}}
\label{eq3bb}
\end{equation}

We use this interaction to calculate self-energies $\Sigma_{00}^<$
and $\Sigma_{00}^>$.

C. The linear response calculation  involves excitations near the
fermi-surface i.e. matrix-elements $K(q,p\sim p_F)$ for which the
$p$-independent interaction (\ref{eq3bb}) is too strong, leading to
divergences if used for calculating the $(10)$ field components.
This is related to a too large numerical value of the related Landau parameter.
Numerical results shown below are therefore made introducing a strength factor $a_{10}$
to define an interaction
$V_a(q)=a_{10}V(q)$. The majority of the calculations presented below
are with $a_{10}=0.3$. Other values are chosen below together with
a discussion of the Landau parameter.  (See section 4)

\subsection{Equilibrium Temperature}

In a KB-calculation with selfenergies  defined by a conserving approximation
the total energy is conserved; while the  potential energy decreases and
kinetic energy increases with the same amount until internal equilibrium.
The result is an increase in temperature from the initially set value.
This situation can be moderated by an imaginary time-stepping method.
\cite{dan84,hsk95,hsk01}
The final temperature $T_f$ in equilibrium will be a function of the initial 
imaginary time $\tau$ (and initial temperature) with $T_f \rightarrow 0$ in the
limit $\tau \rightarrow \infty$.

The temperature $T_f = 1 / \beta$ and chemical potential $\mu$ are then related 
to the equilibrium self-energies $\Sigma^{<}$ and $\Sigma^{>}$ by  \cite{kad62}:
\begin{eqnarray}
\Sigma^{>}({\bf
p},\omega)=-e^{\beta(\omega-\mu)}\Sigma^{<}({\bf p},\omega).
\label{equil}
\end{eqnarray}
(The ratio of the selfenergies is consequently independent
of momentum ${\bf p}$, which serves as a check on numerical accuracy.)

With $\beta$ and $\mu$ calculated from eq. (\ref{equil})  the equilibrium
uncorrelated distribution function is given by
\begin{eqnarray}
f(p)=1/(1+e^{\beta(\omega(p)-\mu)})
\label{ferm}
\end{eqnarray}
with
\begin{eqnarray}
\omega(p)=p^{2}/2m+Re\Sigma^{+}({\bf p},\omega)+\Sigma_{00}^{HF}(\bf(p))
\label{omega}
\end{eqnarray}

The real part of $\Sigma^{+}$ is calculated from the imaginary part by
the dispersion relation, and using the relation

\begin{eqnarray}
2Im\Sigma({\bf p},\omega)=i(\Sigma^{<}({\bf p},\omega)-
\Sigma^{>}({\bf p},\omega)
\label{imsig}
\end{eqnarray}

While the uncorrelated distribution is given by  eq. (\ref{ferm}) the
correlated one is given by
\begin{equation}
\rho({\bf p},t)=-iG^{<}({\bf p},t,t).
\label{corrdis}
\end{equation}

\section{Numerical Results}

Calculations are made for symmetric nuclear matter at normal nuclear matter
density; $\rho=0.13 {\rm fm}^{-3}$, i.e. fermi-momentum $k_F=1.25 {\rm fm}^{-1}$. The
external momentum transfer is chosen to be either $q_0=0.2, 0.4$ or $0.8
{\rm fm}^{-1}$.
The calculations follow essentially the computing-methods described in ref.
\cite{hsk99} with the additional requirement to also time-evolve
$G_{10}$. The eq. \ref{eq4a}   is solved as  a matrix equation.

All numerical results  shown below are without  the initial imaginary
time-stepping referred to above.    The initial (t=t'=0) distribution
$\rho(p)=-iG^<_{00}(p,0,0)$ is taken to be a zero temperature
fermi-distribution except in the uncorrelated case where $T=5$ MeV.(See
below.)
The system is then time-evolved and correlations are completed at
$t\geq t_c$, with $t_c$ being the correlation time. 
In the linear response evolution used here, separating the Green's
function into $G_{00}$ and $G_{10}$ components, the collision term for
the $G_{00}(p,t,t)$ evolution will vanish for $t>t_c$ and  the distribution 
$\rho(p)$ is then stationary.  (See ref. \cite{hsk01} for a detailed discussion of
correlation times $t_c$.)

\begin{figure}
\begin{center}
\includegraphics[width=0.32\textwidth,angle=-90]{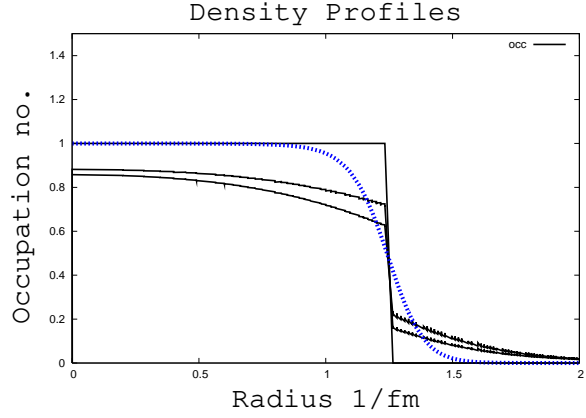}
\caption{
Solid lines show  the  zero temperature Fermi-distribution (top curve), the correlated
distribution with the second-order selfenergy calculations (second from
top) while the bottom curve is with RPA selfenergys. The broken curve (blue on-line)
is a $T=5$  MeV Fermi-distribution.
}
\label{dens}
\end{center}
\end{figure}

Fig. \ref{dens} shows the relevant distribution-functions.
Solid lines  show the initial zero-temerature Fermi-distribution
and two correlated distributions corresponding to the three cases I,II,and III.   
The temperatures calculated from eq. (\ref{equil} are in cases II and III  
$T_f=5.4$ and $5.5$ MeV respectively, while in case I the temperature 
is $T_f=T_i=.0.$  The broken curve  shows the $T=5$ MeV  Fermi-distribution.

The result of the correlations is a depletion of occupation below the
fermimomentum and scattering into states above this mommentum. It is
very similar to results obtained with more 'realistic'
interactions/cite{hsk92}. We  therefore claim that the correlations
induced with our simple model-interaction are 'realistic'.
With the correlations completed in the $(00)$ channel the
external field $U_0(t)$ is now applied generating the $G_{10}$ field
as shown by eqs. (\ref{eq01a}) and (\ref{eq01b}).
Note that with the
conserving approximation satisfied the vertex-correction is
'automatically' included in the ensuing time-evolution.\cite{nhk00}

\begin{figure}
\begin{center}
\includegraphics[width=0.32\textwidth,angle=-90]{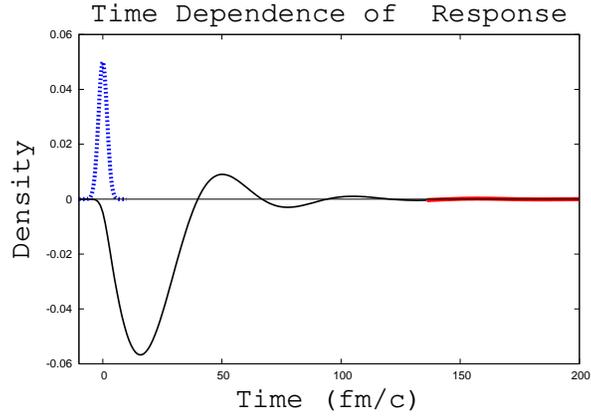}
\caption{
The solid curve  shows the density  $\delta n(q_0,t)$ with $q_0=0.4 {\rm fm}^{-1}$ 
for $t<120 fm/c$ and extrapolated for $t>120 fm/c$ (red on-line). The broken  
curve (blue on-line) shows the external field $U(q_0,t)$ with a maximum of 
$50$ MeV at $t=0$.
}
\label{time}
\end{center}
\end{figure}

The  resulting time dependent  density $$\delta
n(q_0,t)=-i\sum_{\bf p} G^{<}_{10}({\bf p},t,t)$$ is recorded
until it is fully damped.
If the damping time is too large (number of time-steps $> \sim 150$)
then $n(q_0,t)$ is extrapolated using the amplitudes and frequencies
of the oscillations for lesser times.
A typical result of the evolution in real time
is shown in Fig. \ref{time}.
This time-function is then fourier-transformed to $\omega$-space.
Results for the three different examples of self-energies
referred to in the Introduction are shown below.


\subsection{Mean field +RPA; case I}
All selfenergies except the HF are here identically zero.  Results of these 
calculations are shown by the broken lines (blue on-line) in Fig. \ref{resp3}.
(For comparison with the correlated cases (II and III) the temperature
is here chosen to be $5$ MeV).

There is an overall agreement with  previous RPA calculations e.g.
\cite{boz05,gar92,gog77}.  There are
some differences to be expected because of the variety of  two-body
interactions and mean fields that have been used. 
Our results here are displayed mostly for comparison with the results
below where correlations beyond HF are considered.
The heating to $T=5$ MeV  causes only a very slight broadening in
comparison with a $T=0$ calculation.


\subsection {$2^{nd}$ order correlations; case II}
The Green's functions are in this section dressed by second order
insertions but no rings.  
This means that only the last term in eq. (\ref{eq1}) is kept here i.e.:
$$V_{s}^{^{>}_{<}}({\bf p},t,t') =
V^{2}({\bf p})\Pi_{00}^{^{>}_{<}}({\bf p},t,t')$$
and
$$V_{s(10)}^{^{>}_{<}}({\bf p},t,t') =
V_a^{2}({\bf p})\Pi_{10}^{^{>}_{<}}({\bf p},t,t')$$
\begin{figure}
\begin{center}
\includegraphics[width=0.32\textwidth,angle=-90]{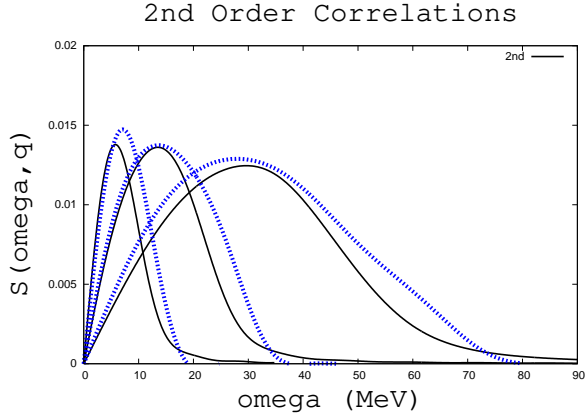}
\caption{
Solid curves show response functions $S(\omega,q_0)$ with $2^{nd}$
order correlations (case II) included.  External momentum transfers are $q_0=0.2,
0.4$ and $0.8 {\rm fm}^{-1}$ respectively. They are compared
with HF+RPA (case I) results at
temperatures  T=5 MeV: broken curves (blue online). See text for further
explanations.
}
\label{resp3}
\end{center}
\end{figure}

The initial temperature
at time $t=0$ is here  equal to zero;
the momentum-distribution is that of a zero-temperature fermi-distribution. The
correlations result in a heating to $T=5$ MeV and an associated smoothing of the
fermi-surface. An additional and more significant smoothing is due to the
broadening of the static spectral-functions in the correlated medium, related to the
imaginary parts of the self-energies also resulting in
a considerable depletion of the distribution at low momenta.(See Fig. \ref{dens})
This is the most probable cause for the difference between the correlated and
uncorrelated cases in the region of the large values of frequencies $\omega$ as
shown in Fig. \ref{resp3}.
Other than that there is in fact no other major effect of the correlations.

\subsection{Full Ring-correlations; case III}
\begin{figure}
\begin{center}
\includegraphics[width=0.32\textwidth,angle=-90]{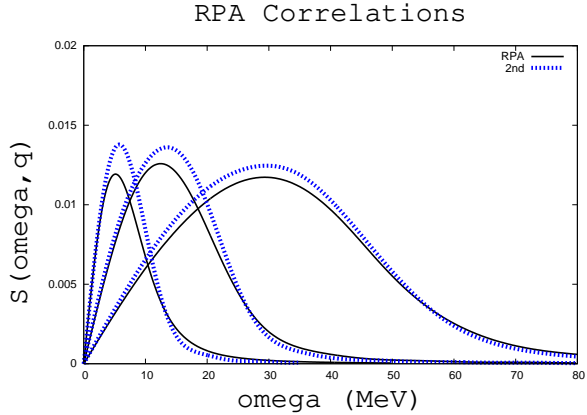}
\caption{
Solid curves show response functions $S(\omega,q_0)$ with full ring correlations (RPA)
included and for external momentum transfers $q_0=0.2, 0.4 $ and $0.8 {\rm fm}^{-1}$
respectively. They are compared with $2^{nd}$ order results:  broken curves (blue online).
See text for further explanations.
}
\label{resp4}
\end{center}
\end{figure}
The results above, showing the effect of including correlations to
second order in the chosen interaction is here extended by also 
including  polarisation bubbles (rings) to all orders. The relevant additional terms are
included in   eqs. \ref{eq1}
and \ref{eq4a}. 
This seems to be  a logical
extension of the second order calculation which just includes  the lowest order
of the full ring expansion. Rings to all orders are of course already generated in
$G_{10}$ by the external disturbance $U(q_0,t)$, yielding the collective oscillations.
Fig. \ref{resp4} shows our result. The $2^{nd}$ order results included in Fig.
\ref{resp3} are for comparison also  shown here
by the broken (blue) lines. The effect is  mainly  that of some  additional  
smoothing with a shift in the distribution from low to  higher  frequencies.
This is related to the additional effect of correlations on the distribution
functions shown in Fig. \ref{dens}.

\subsection{Mean field and energy sum rule}

\begin{figure}
\begin{center}
\includegraphics[width=0.32\textwidth,angle=-90]{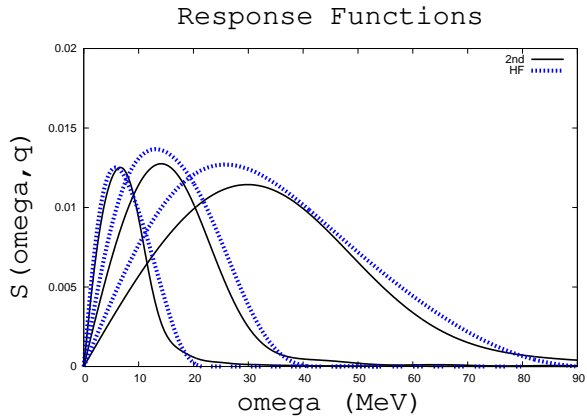}
\caption{
The second order (case II, solid lines) and the un-correlated case
(case I, broken lines, blue on-line) response functions are 
shown here, both with the same mean field, defined by an effective mass
 $m^*=0.7m$, implying $\Sigma^{HF}_{00}(p)\sim p^2$ . 
}
\label{effect}
\end{center}
\end{figure}

Fig. \ref{dens} shows that the density-profiles $\rho(p)$
and therefore also the mean fields calculated from eq. (\ref{VG}) are
different in each of the three cases  that we considered.
To see the effect this might have on our results shown so far
we substituded the mean field $\Sigma^{HF}_{00}$ by an effective mass $m^*=0.7$.
The result is shown in Fig. \ref{effect}.
Compared with Fig.
\ref{resp3} one finds that the different mean fields have some effect but the
differences at the high frequency tails of the response functions  remain.
This calculation also provides a test of the energy sum-rule,
\begin{equation}
\int_{0}^{\infty}\omega S(\omega,q_0)d\omega =\frac{q_0^2}{2m^*}
\label{EE}
\end{equation}
that cannot be directly validated with the mean field given by eq.
(\ref{VG}) because
it does not have a simple $p^2$ dependence. (The effective mass
$m^*$ is $p$-dependent; $m^*(p)=\frac{p^2}{2m\Sigma^{HF}(p)+p^2}.$) With
vertex corrections 'automatically'
included in our 2-time formalism (see ref. \cite{nhk00}) this test is of no
issue in the present work. 
It is in our calculations  more a test of numerical consistency than a test of
the formalism and this sum-rule was found to be satisfied within numerical accuracy.

\subsection{Density Fluctuations}
\begin{figure}
\begin{center}
\includegraphics[width=0.32\textwidth,angle=-90]{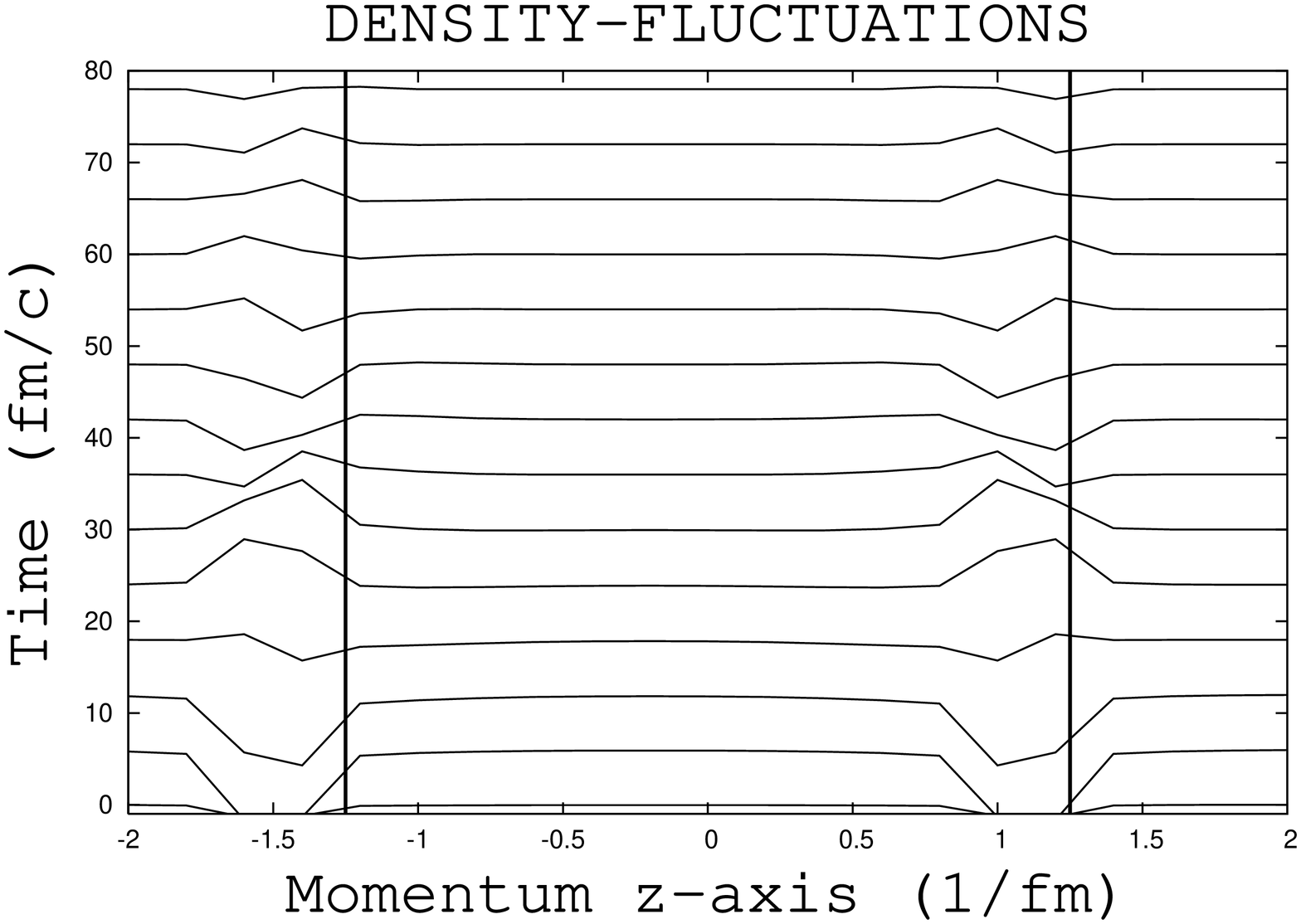}
\includegraphics[width=0.32\textwidth,angle=-90]{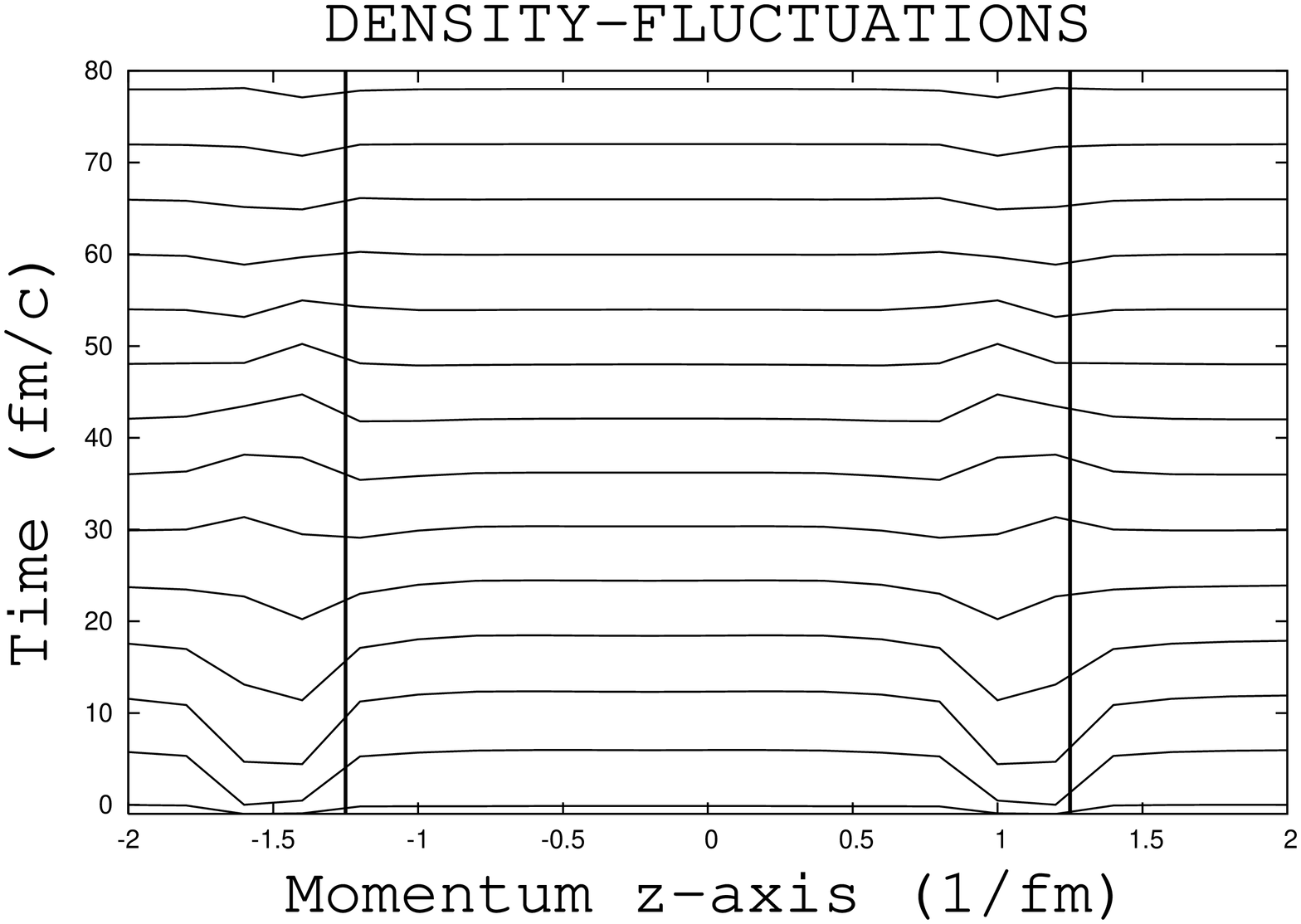}
\caption{
The perturbed density integrated over perpendicular momenta  is shown as
a function of momentum $p_z$ (with ${\bf q_0}$ along the z-axis) and time
$t$; to the left with selfenergies are here calculated to second order
and to the right .
with selfenergies
calculated to all orders of polarisation bubbles, i.e. RPA.
The vertical lines (blue online) are at $p_z=\pm p_F$ (fermi-momentum).
}
\label{fluct}
\end{center}
\end{figure}

In order to show the approach to equilibrium in some detail the density
$iG^<_{10}({\bf p},t,t)$ was  integrated over $p_x$ and $p_y$ ,
($p_z$ being the component along ${\bf q_0}$). The resulting $t$ and $p_z$
dependent functions are shown
in Fig. \ref{fluct}  for $q_0=0.4 {\rm fm}^{-1}$ for selfenergies
calculated to second (left figure) and all orders (right figure).
One finds that the equilibration following the external perturbation is noticably
faster when including the rings to all orders (full RPA)
in the self-energies than  to $2^{nd}$ order only.

This difference in damping between the two cases does not seem so obvious when
comparing the corresponding response functions shown in Fig. \ref{resp4}, although 
a slight difference in the widths of the distributions can be noticed.

As already concluded above, 
the main effect of the correlations is to be found at the high frequency end of
the response function. In the long wave-length limit the response function is
mainly dependent on the excitations close to the fermi-surface shown by 
Fig. \ref{fluct}. 
The correlations result in states populated above the fermi-momentum that when excited
by the external potential contribute to the energy spectrum at the high end of the
response function. 

\newpage
\section{Landau parameters}
Owing to the relatively low temperature and long fluctuation wavelengths
considered here, the excitations in our linear response calculations are concentrated
around the fermi-surface. This is clearly evident from  Fig.
\ref{fluct}   which, as pointed out above, can explain why
the  effect of the correlations have a smaller effect  than might be expected,
repopulating the states close to the fermi-surface.
This is  also why   Landau-theory would be  applicable here as the quasi-particle
picture dominates scatterings close to the fermi-surface. Another related reason why
this theory is pertinent to our work is the  relation
\cite{bab73} between the dimensionless Landau-parameter
\begin{equation}
F_0=\frac{2m^*k_F}{\pi^2\hbar^2}f_0
\label{lnd}
\end{equation}
and the summation of ring-diagram correlations,  the collective excitation we are
dealing with here. The factor $f_0$ is here the angular momentum ($l=0$)
quasi-particle (effective) interaction.
This Landau parameter relates to the compressibility $K$ by
\begin{equation}
K=6e_F(1+F_0)
\label{K}
\end{equation}

The interaction $f_0$ is composed of  a 'direct' part defined by the functional 
derivative $$f_{{\bf kk'}}=\frac{\partial^2E}{\partial n_{{\bf k}}n_{{\bf k'}}}$$ 
(with ${\bf k} \rightarrow{\bf k'}$) and an 'induced' part.

The 'induced' interaction relates to our discussion above as it 
is essentially the effect of screening by the
ring-diagrams. It is  defined below. Screening was first included in 
nuclear shell model calculations by Bertsch by a second order 'core-polarisation' 
diagram.\cite{ber65} and later calculated to all orders by Kirson.\cite{kir71}

Guided by methods used by Babu and Brown \cite{bab73}, who considered interactions 
in liquid $^3He$, Sj\"oberg \cite{sjo73} and  later others (see e.g. ref. \cite{gam11})
calculated Landau parameters  for nuclear matter.  They used Brueckner's $K$-matrix  
expression for the total energy $E$ in their analysis.  The 'direct' interaction is 
then just the Brueckner $K$-matrix while the 'induced', $f_i$  is of higher orders of $K$.
They find  it to be given   by $$f_i({{\bf k,k}})=
-\sum_{{\bf p,p'}}f({\bf k,p})\left(\frac{\delta n_{{\bf p}}({\bf q},\omega)}
{\delta U_{{\bf p'}}({\bf q},\omega)}\right)_{q,0}f({\bf p'k})$$

Note that the perturbative potential $U_({\bf p'})$ is here assumed to act on
particles of momenta ${\bf p'}$ only, to generate density fluctuation 
$\delta n_{{\bf p}}$.  Our interactions are not 'realistic' enough to warrant a detailed
comparison with Sj\"oberg's (or other's based on Brueckner theory).
It is here made only to illustrate the connection of our work with that
based on Brueckner theory.\cite{bab73,sjo73}
For an  approximate comparison we calculate the 'induced' interaction  by 
$$f_i=V_a^2(0)\frac{\delta n_{max}}{U_{max}}$$,where as shown in 
Fig. \ref{time} $U_{max}=50$ MeV and $\delta n_{max}$ is the maximum in the 
density fluctuation.
From the result of  the calculation of response function shown in Fig.  \ref{resp3}
(2nd order correlations) for $a_{10}=.3$ and for $q=0.2 {\rm fm}^{-1}$
we find $\frac{\delta n}{\delta U}=.00126$ to get for the 'induced'
interaction $24$ Mev${\rm fm^3}$  with the 'direct' being $-140$  MeV ${\rm fm^3}$ to get
$F_0=-0.495$.  For comparison  Sj\"oberg obtained  $F_0=-0.373$ for the
angular momentum l=0 state.
When comparing the two results one has to remember that Landau theory is a 
quasi-particle theory while our calculation goes beyond the quasi-particle with
complex self-enrgies and non-zero spectral widths.
Also, our result is with $q_0=0.2 {\rm fm}^{-1}$ not $q_0=0$ which is not dirctly
attainable with our method.

\begin{figure}
\begin{center}
\includegraphics[width=0.32\textwidth,angle=-90]{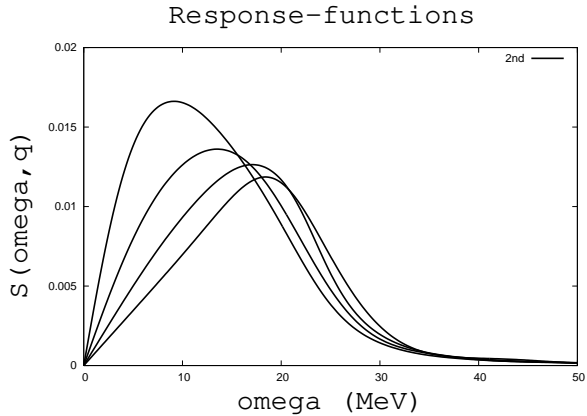}
\caption{
Solid curves  show  response functions $S(\omega,q_0=0.4)$ with
second order self-energies and for $a_{10}=0.4,0.3,0.2$ and $0.1$ respectively
from left to right. An effective mass $m^*=0.7$ was used here. 
}
\label{land}
\end{center}
\end{figure}

The comparison of our calculated value of $F_0$ with that of Sj\"oberg's suggests
however that our choice $a_{10}=0.3$ is somewhat too large. 
Table 1 shows the dependence of $F_0$ on the strength parameter $a_{10}$ and 
Fig. \ref{land}  shows the response function $S(\omega,q_0=0.4 {\rm fm^{-1}}) $
for $a_{10}=.1,.2,.3$ and $0.4$.
\begin{table}
\begin{center}
TABLE 
\end{center}
\centerline{
  \begin{tabular}[t]{|c |c |c |c | c |}
  \hline
  & $ a_{10} $ & $f_0^{(d)}   $ &  $f_0^{(i)}  $    &  $F_0  $
  \\ \hline
  &  0.1       &  -47.0         &  2.0              & -0.190 
  \\ \hline
  &0.2         &  -93.0         &  10.0             & -0.356
  \\ \hline
  &0.3         &  -140.0        &  24.0             & -0.495
  \\ \hline
  &0.4         &  -187.0        &  46.0             & - 0.602
  \\ \hline
   \end{tabular}
}   
   \end{table}

The result is as to be expected; a considerable shift of the distribution with the
strength. The stronger interaction excites states deeper in the fermi-sea while
the weaker only excites states at the edge of the distribution i.e. of higher
energy.

We showed how the Landau parameters can be calculated from our response function
calculations. 
Alternatively, the response-function may be calculated using interactions given by
Landau parameters.
There are several published results using this method. (see ref. \cite{pas14} and
references therein). 

\section{Summary and Conclusions}
The main purpose of this presentation is to illustrate the usefulness
of the two-time Green's function (Kadanoff-Baym) method for the study
of nuclear response. It was previously applied to the electron gas to
see the effect of correlations.\cite{nhk00}  The time-evolution of
Green's function in two-time space with conserving approximations for
the self-energies guarantees the preservation of the energy sum-rule.
Alternative methods (in $\omega$-space) has been reported by Bozek 
et al.\cite{boz04,boz05} and  by Lykasov et al.\cite{lyk04}.

It may be a matter of choice but it seems that the two-time method has
some advantages numerically.  Calculations at zero temperature and/or with small
values of $q_0$ for which the spectral width is
small, are for instance a problem  when working in $\omega$-space, but less so in the
2-time calculation. 

The rise in temperature associated with the onset of correlations when
using the two-time method is sometimes cited as a drawback. It is
however not difficult to remedy by allowing the correlations to
initially form by propagating along the  imaginary time-axis. 
This was not done here but left for future works.

Comparing the various results obtained so far in our on-going investigation
one finds some  differences between the three different
approximations (collisionless, $2^{nd}$-order and full rings)
at high frequencies with corresponding decreases (but without any
significant shift) in peak energies.
In this regard our result agrees with those of ref. \cite{boz05}.
Whether the effect of correlations on e.g. neutrino-absorption or other medium
sensitive reactions is important or not remains to be investigated.

The method of the two-time Green's function time-evolution is well
documented, but several improvements in its application to the problem
of response remain, mainly  problems associated with any
nuclear many body calculation i.e. that of the NN-force itself as well as
in-medium effects.

We here circumvented this problem by defining three different
potential-strengths, $V_G(p), V(q)$ and $V_a(q)$ relating to the mean
field, correlations and external field excitations respectively. This
involved admittedly a rather \it ad hoc \rm choice but acceptable for this
preliminary investigation.
The most important choice of these  is that of $V_a(q_0)$ entering in the
calculation of the Hartree field that is the driving mechanism of the response.
The bulk of the numerical results shown above were all made with the
parameter $a_{10}=0.3$ defining the strength of $V_a$.
The choice of this particular value for this strength-factor was not based on any 
'state of the art' calculation of the in-medium effective interaction. It  was only chosen
as to provide a reasonable representation of the interactions at the fermi-surface
It  led to a reasonable value of the related Landau-parameter. We also showed the 
dependence of this parameter  on the strength-factor $a_{10}$ as well as the corresponding
response functionsi as shown in Fig. \ref{land}.

Spin,isospin, tensor,spin-orbit and pairing are however also aspects of 
the interaction that are of interest to
consider\cite{boz05,lyk04,dab76,ols04,pas14,mar06,sed10}
but still remain to be included in our calculations where we only
considered density fluctuations.
It has in addition been shown that the non-central parts of the effective
interactions are modified close to the fermi-surface.\cite{sch04} This may in 
particular effect spin-excitations in the long-wave length limit with
excitations  close to the fermi-momentum.

Although our understanding of nuclear forces and of the nuclear many-body
problem is under constant development it is still incomplete.  There is
however important present-day  knowledge
that can and should  be incorporated to improve  our nuclear response
calculations.

One reason for choosing a local potential (eq. \ref{ggg}) as done here
is simply that the 2-time computer-code
(\cite{hsk99}) used here is restricted to this choice.
Separable interactions have however been obtained from inverse scattering. They are
'realistic' in the sense that they by construction fit scattering and deuteron data. 
\cite{kwo97} They can now be incorporated in our calculations  by  a new 2-time 
code (\cite{hsk15})  developped  specifically for these  separable (non-local) potentials.

The results shown are for symmetric nuclear matter. Response calculations
for neutron matter is of particular interest related to astrophysical
problems, in particular neutrino-absorption.

As a further comment we like to point out that
the self-energy insertions have been separated into a Hartree-Fock (mean
field)
that is real and a 'correlation' part which is complex. This can be done
consistently in the case of weak interactions. With medium-dependent
effective interactions such a separation is not well defined and can
lead to double-counting.
The real term of the 'correlation' part
adds to the mean field but if derived from an effective interaction it would 
already be included in the  mean field {\it if this mean field is derived from the same
effective interaction}.
So one may claim that a double-counting can  occurr.
We avoided this situation here by choosing the three different forms of interaction
referred to above.

\section{Appendix:Diagrammatic representation of linear response in the
real-time non-equilibrium Green's function formalism}
\label{formalism.sec}
\begin{figure}[ht]
\centerline{\includegraphics[scale=0.6,angle=0,trim=50 220 00  90,clip=true]
{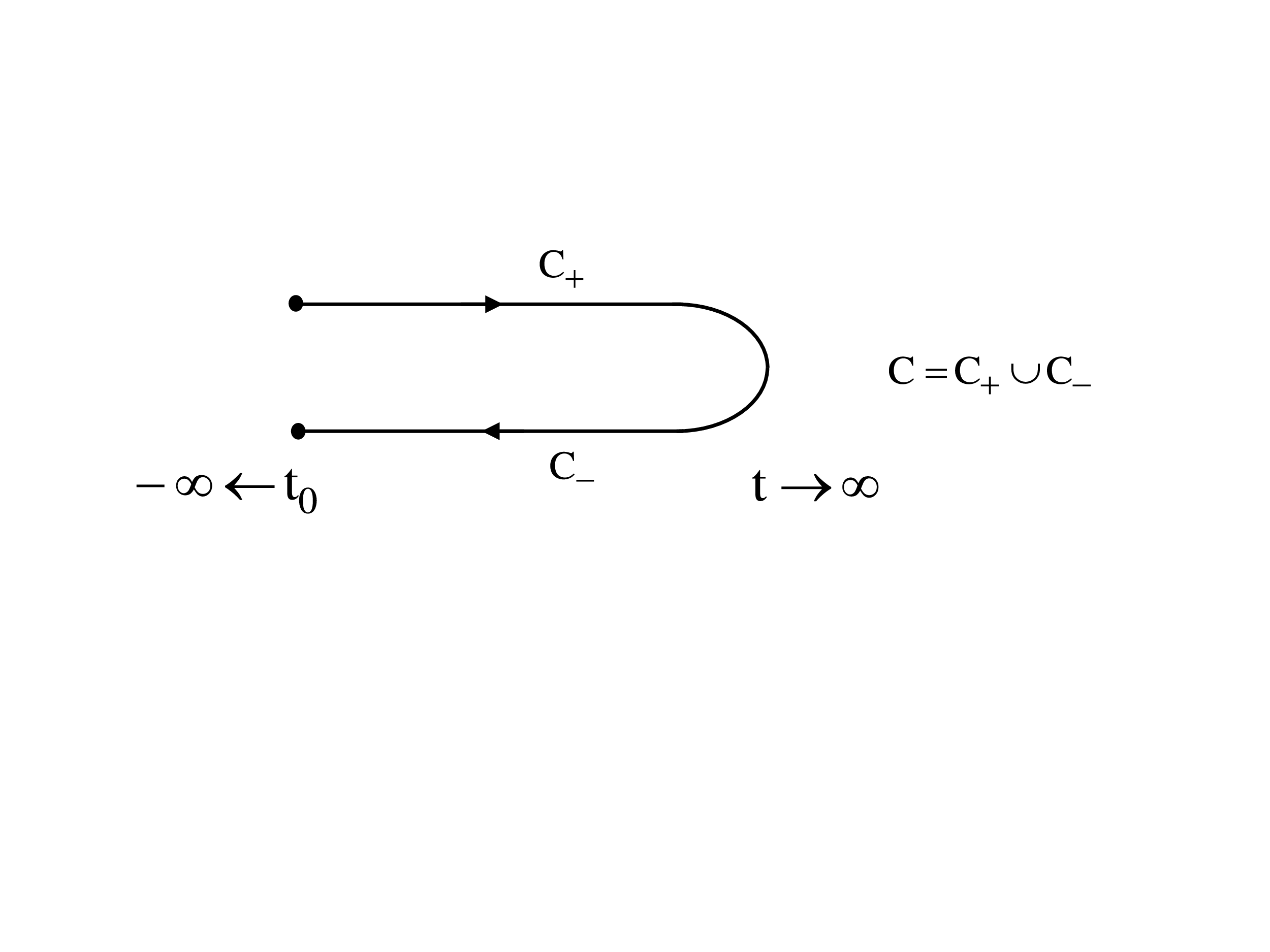}} 
\caption{The double-time contour on which the non-equilibrium Green's functions are defined.
The branch $C_+$ goes from a reference time $t_0$ in the distant past to $t \rightarrow \infty$, 
and $C_{-}$ goes from $\infty$ back to $t_0$. 
} 
\label{Keldysh-contour.fig} 
\end{figure}

\begin{figure}[ht]
\centerline{\includegraphics[scale=0.5,angle=0,trim=00 00 00 74 00,clip=true]
{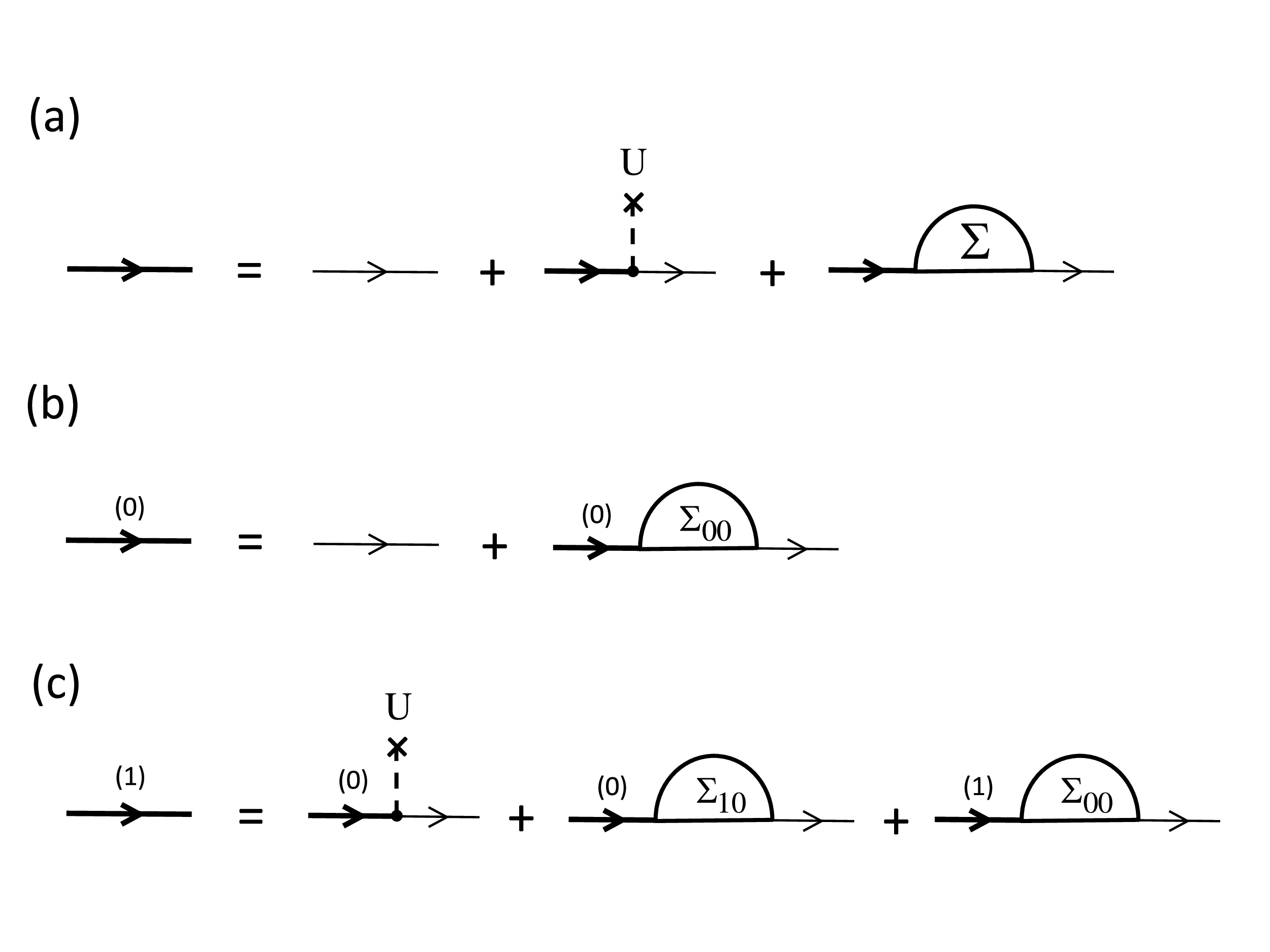}}
\caption{Dyson equation for the nonequilibrium single-nucleon Green's function. The
thick solid directed line, the thin directed line, the dashed line labeled $U$, 
and the semicircle labeled $\Sigma$ denote, respectively, the Green's
function, the free-nucleon Green's function, the external     potential, and the self-energy. 
The number in parentheses, (0) or (1), above a Green's function line gives the order 
of $U$ in that Green's function. $\Sigma_{00}$ ($\Sigma_{10}$) is that part 
of $\Sigma$ evaluated to zeroth (first) order in $U$. }
\label{dyson.fig}
\end{figure}

\begin{figure}[ht] 
\centerline{\includegraphics[scale=0.5,angle=0,trim=00 00 00 50,clip=true]
{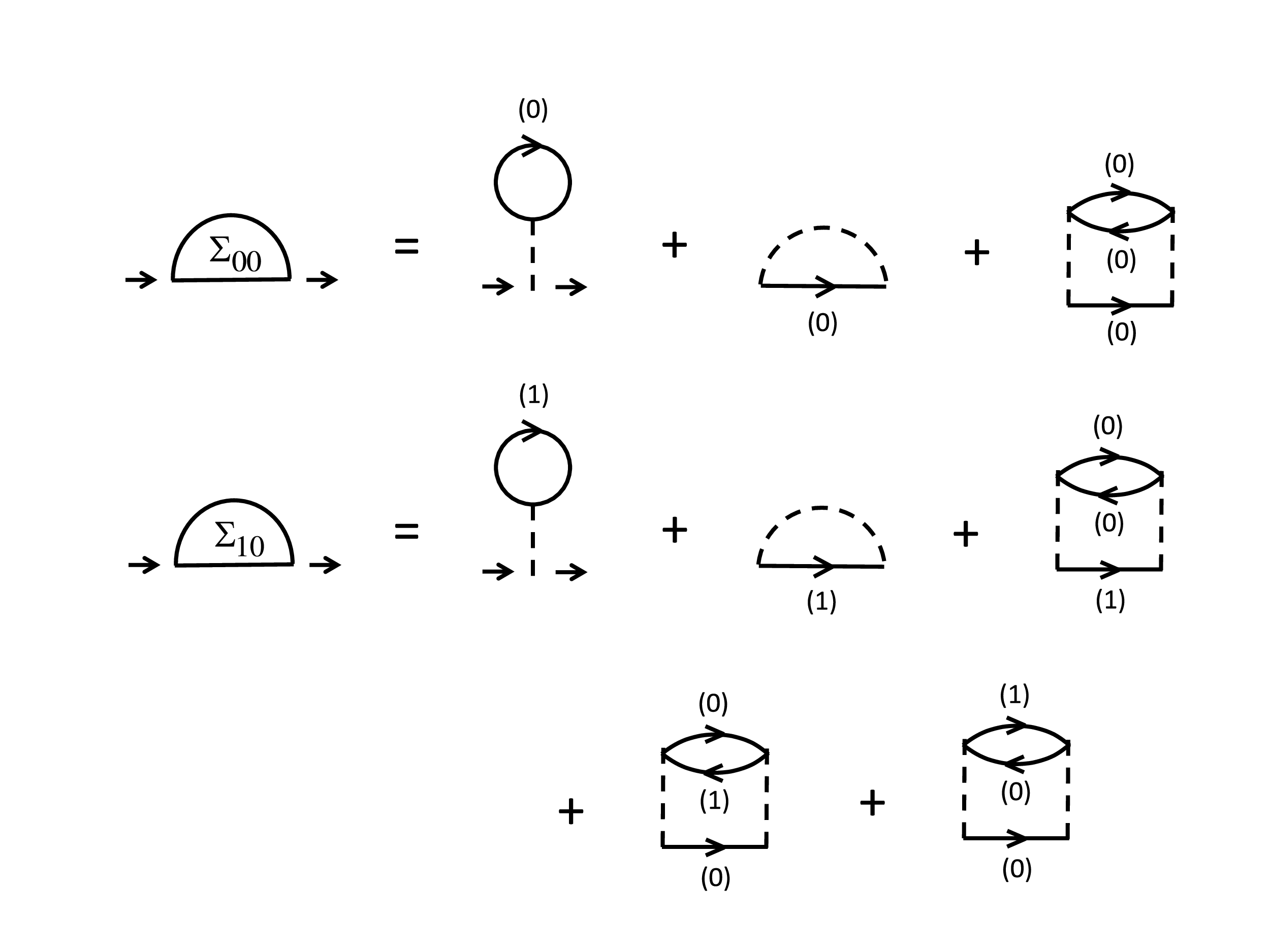}} 
\caption{Zeroth and
first order (in $U$) self energy in second order Born approximation. All
solid lines are interacting Green's functions, and the dashed line denotes
the nucleon-nucleon intera    ction. In each equation, retaining only the first
two terms yields the Hartree-Fock approximation.}
\label{Sigma-2nd-order.fig} 
\end{figure}

\begin{figure}[ht] 
\centerline{\includegraphics[scale=0.5,angle=0,trim=00 00 00 70,clip=true]
{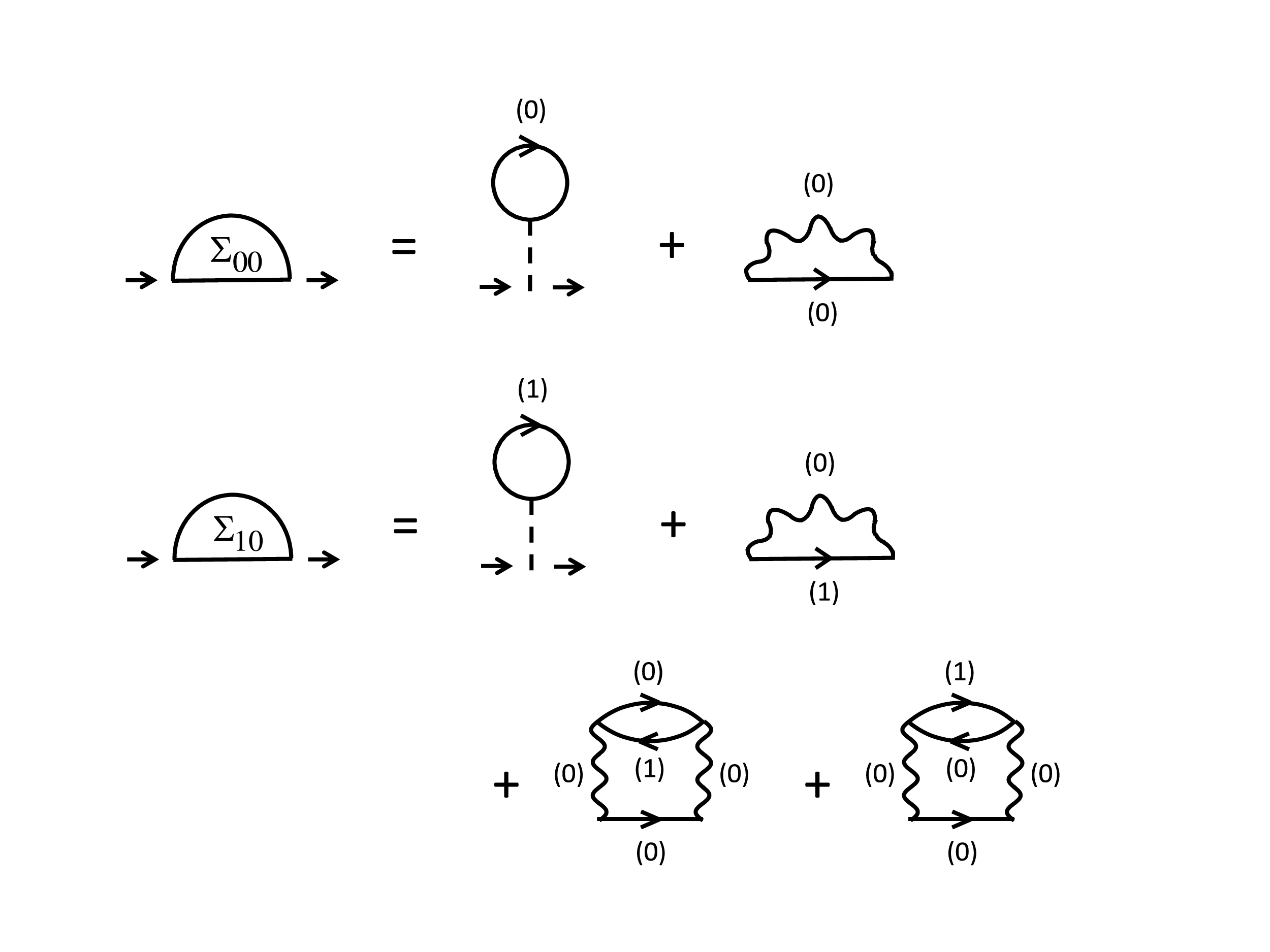}}
\caption{Self energy in the ring-diagram approximation. The wavy line denotes 
the nucleon-nucleon interaction dressed by ring diagrams which is shown in Fig.
\ref{screened-potential.fig}.}
\label{Sigma-ring-diagrams.fig}
\end{figure}

\begin{figure}[ht]
\centerline{\includegraphics[scale=0.5,angle=0,trim=00 250 00 00,clip=true]
{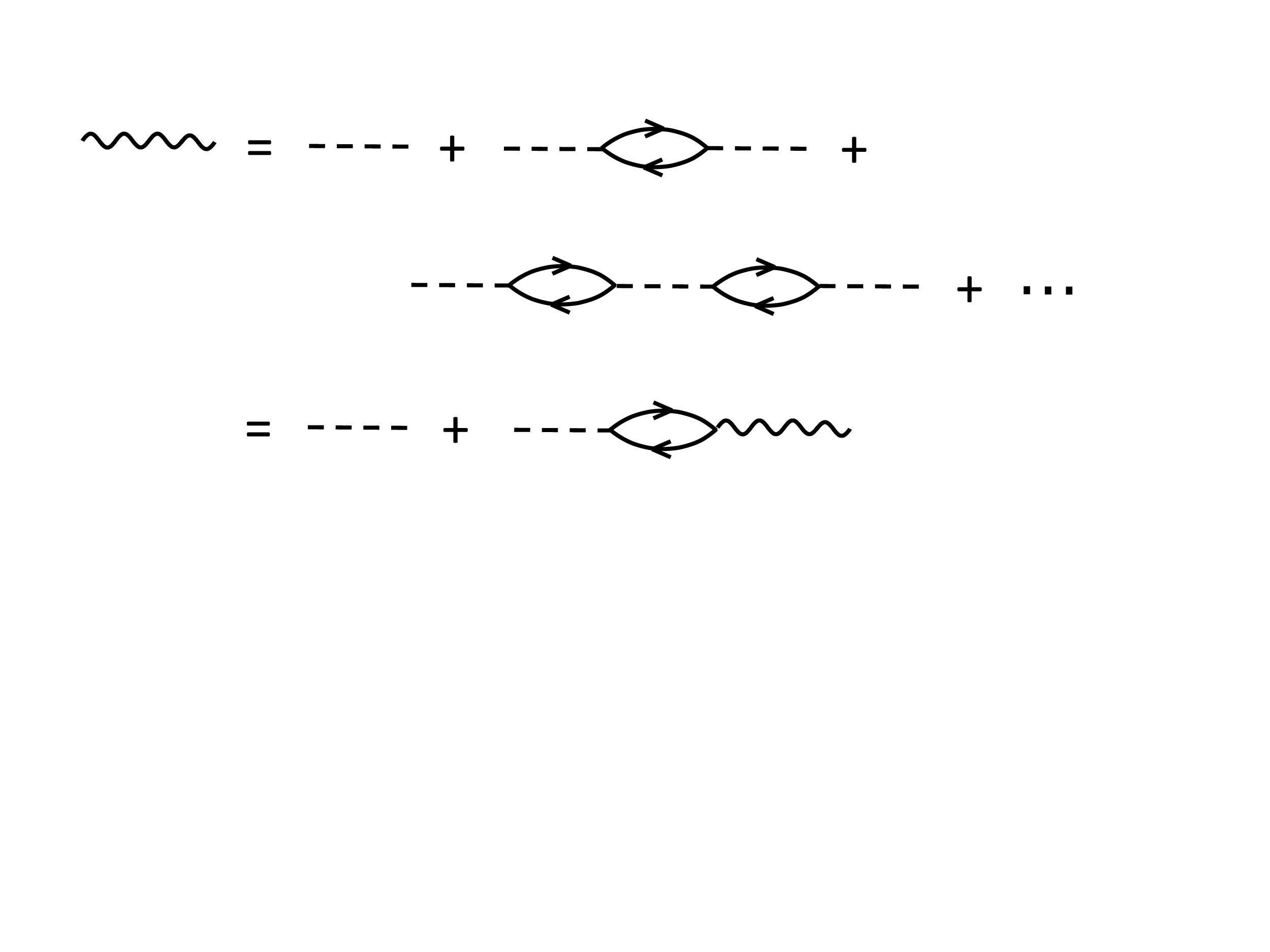}}
\caption{Nucleon-nucleon interaction dressed by ring diagrams. Note that
the solid lines are fully  dressed Green's functions.}
\label{screened-potential.fig}
\end{figure}

In this appendix, we summarize the linear response of the
many-nucleon system to an external field in the real-time
Green's function formalism \cite{keldysh.65,fujita.65,craig.68,danielewicz.84, 
kad62, botermans-malfliet.90,kremp-etal.05}.
 We use the standard diagrammatic
perturbation formulation of Green's functions on a double-time
contour . The diagrammatic representation facilitates
comparison with the equivalent finite-temperature equilibrium
Green's function treatment when the zeroth-order (in the
external field) many-body state is an equilibrium state. We
note that the linear response theory as formulated here is more
general than that based on equilibrium Green's functions. Our
method of solution is valid when the zeroth-order state is itself
evolving in time. Baym-Kadanoff's papers.
The Green's functions are defined with time arguments
residing on a double-time contour shown in  Fig.
\ref{Keldysh-contour.fig}. The single-nucleon basis orbitals are
labeled by momentum ${\bf k}$, spin projection $\sigma = \pm
\frac {1} {2}$, and isospin projection $\tau = \pm \frac {1} {2 }$. 
The single-nucleon Green's function is  defined as
\begin{equation} 
G \left( {\bf k}_1 \sigma_1 \tau_1 \bar{t}_1 , {\bf k}_2 \sigma_2 \tau_2
\bar{t}_2 \right) = - i \langle T_C [ a_{{\bf k}_1 \sigma_1 \tau_1}
( \bar{t}_1 ) a_{{\bf k}_2 \sigma_2 \tau_2}^{\dagger} ( \bar{t}_2 ) ]
\rangle
\end{equation}
where $a_{{\bf k} \sigma \tau} ( \bar{t} )$ and
$a_{{\bf k} \sigma \tau}^{\dagger} ( \bar{t} )$ respectively annihilates
and creates a nucleon in state (${\bf k} \sigma \tau$) at time
$\bar{t}$.
In this paper, a bar over a time symbol indicates that the
represented time resides on the double-time contour, denoted by
$C$, in Fig. \ref{Keldysh-contour.fig} while an unbarred
time symbol represents a point on the regular time axis.
$T_C$ represents time-ordering along the time contour $C$, and
$\langle \cdots \rangle$ denotes taking an expectation value
in the initial state of the system. The time ${\bar t}$ on $C$
can also be written as $(t , b)$ where $t$ is the regular
time and $b = + (-)$ labels the branch $C_{+}$ ( $C_{-}$ )
that ${\bar t}$ is on. The Green's function can then be written
in a four-component form 
$G^{b_1 b_2} \left( {\bf k}_1 \sigma_1 \tau_1 t_1 , {\bf k}_2 \sigma_2 \tau_2 t_2
\right)$ where
\begin{eqnarray}
G^{+ -} \left( {\bf k}_1 \sigma_1 \tau_1 t_1 , {\bf k}_2 \sigma_2 \tau_2 t_2 \right) &=& i \langle [ a_{{\bf k}_2 \sigma_2
\tau_2}^{\dagger} ( t_2 ) a_{{\bf k}_1 \sigma_1 \tau_1} ( t_1 ) ] \rangle \\
\nonumber
&\equiv & G^{<} \left( {\bf k}_1 \sigma_1 \tau_1 t_1 , {\bf k}_2 \sigma_2 \tau_2 t_2 \right) \label{G-pm.equ} \\
G^{- + } \left( {\bf k}_1 \sigma_1 \tau_1 t_1 , {\bf k}_2 \sigma_2 \tau_2 t_2 \right) &=& - i \langle [ a_{{\bf k}_1 \sigma_1
\tau_1} ( t_1 ) a_{{\bf k}_2 \sigma_2 \tau_2}^{\dagger} ( t_2 ) ] \rangle \\
\nonumber
&\equiv& G^{>} \left( {\bf k}_1 \sigma_1 \tau_1 t_1 , {\bf k}_2 \sigma_2 \tau_2 t_2 \right) \label{G-mp.equ} \\
G^{+ +} \left( {\bf k}_1 \sigma_1 \tau_1 t_1 , {\bf k}_2 \sigma_2 \tau_2 t_2 \right)
&=& - i \langle T_{+}[ a_{{\bf k}_1\sigma_1 \tau_1}( t_1 ) a_{{\bf k}_2 \sigma_2 \tau_2}^{\dagger}( t_2 )]\rangle
\label{G_pp.equ} \\ \nonumber
&=& \theta ( t_1 - t_2 ) G^{>} \left( {\bf k}_1 \sigma_1 \tau_1 t_1 , {\bf k}_2 \sigma_2 \tau_2 t_2 \right)+ \\
&&  \theta ( t_2 - t_1 ) G^{<} \left( {\bf k}_1 \sigma_1 \tau_1 t_1 , {\bf k}_2 \sigma_2 \tau_2 t_2 \right) \nonumber \\
G^{- -} \left( {\bf k}_1 \sigma_1 \tau_1 t_1 , {\bf k}_2 \sigma_2 \tau_2 t_2 \right) 
&=& - i \langle T_{-}[ a_{{\bf k}_1 \sigma_1 \tau_1}( t_1 ) a_{{\bf k}_2 \sigma_2 \tau_2}^{\dagger}( t_2 )]\rangle 
\label{G_mm.equ} \\ \nonumber
&=& \theta ( t_1 - t_2 ) G^{<} \left( {\bf k}_1 \sigma_1 \tau_1 t_1 , {\bf k}_2 \sigma_2 \tau_2 t_2 \right)+ \\
&&  \theta ( t_2 - t_1 ) G^{>} \left( {\bf k}_1 \sigma_1 \tau_1 t_1 , {\bf k}_2 \sigma_2 \tau_2 t_2 \right) \nonumber
\end{eqnarray}
from which one can see that only two of the four components are independent. 
In Eqs. (\ref{G_pp.equ}) and (\ref{G_mm.equ}), $T_{+}$ denotes time-ordering and
$T_{-}$ anti-time ordering (ordering operators with earlier ones on the left of 
later ones). 
The perturbation theory for calculating the Green's functions and the diagrammatic 
tools that help implementing the theory can be found in e.g.  
\cite{keldysh.65,fujita.65,craig.68,danielewicz.84}.
Grouping classes of diagrams yields equations of motion
and non-perturbative approximation schemes for the Green's functions
\cite{kad62,botermans-malfliet.90,kremp-etal.05}.
We show the diagrammatic content of the approximations we use for the linear response
calculations in this paper.
Fig. \ref{dyson.fig} shows the diagrams representing the equation of motion, or Dyson equation,
of $G \left({\bf k}_1\sigma_1\tau_1t_1,{\bf k}_1\sigma_1\tau_1t_2  \right)$. 
The three parts show, 
(a) the full equation, (b) the equation to zeroth order in the external potential $U$, 
and (c) the equation to first order in $U$. The first order equation is written
in terms of $G^{^{>}_{<}}_{10}$  as Eqs. (1) and (2)  in the main text, where the
spin and isospin labels are omitted in the Green's functions. 
The self energies, to zeroth and first orders in $U$, are evaluated in this paper in various
approximations, which are represented in Figs.  \ref{Sigma-2nd-order.fig} and
\ref{Sigma-ring-diagrams.fig}. The diagrams representing processes up to the second Born
approximation are shown in Fig.  \ref{Sigma-2nd-order.fig}. In each equation in
this figure, the first two diagrams on the right h and side give the Hartree-Fock 
approximation. The processes in the ring-diagram approximation are shown in 
Fig. \ref{Sigma-ring-diagrams.fig}, where the wavy line denotes the in-medium 
nucleon-nucleon interaction dressed by ring diagrams and is given in Fig. 
\ref{screened-potential.fig}. 
		       
 
\begin{figure}[ht]
\centerline{\includegraphics[scale=0.5,angle=0,trim=00 00 00 20,clip=true]
{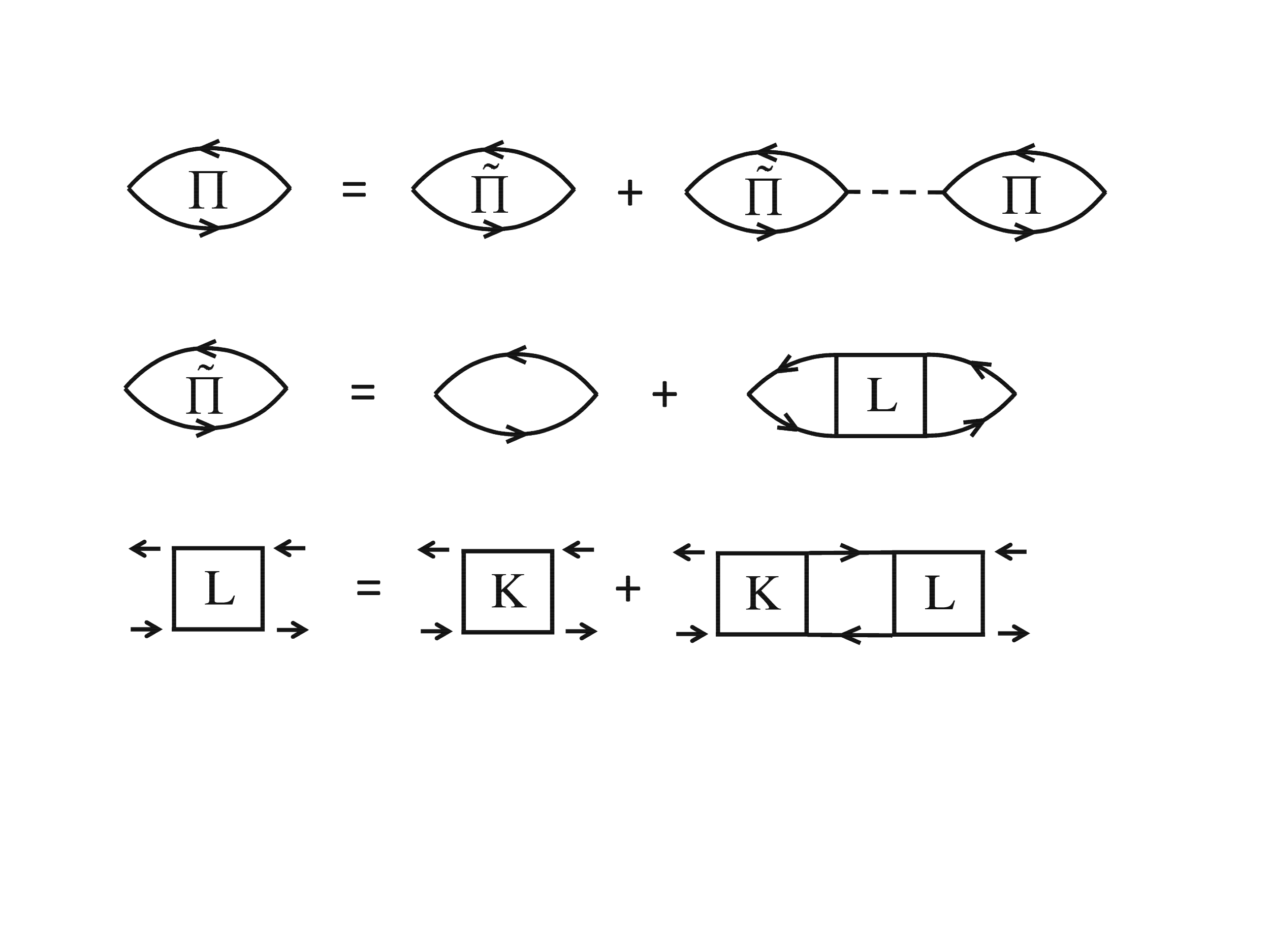}}
\caption{Diagrams representing the density correlation $i \Pi({\bf q},{\bar t},{\bar t}')$
which yields the retarded correlation function in the conventional
formulation of linear response.  All solid lines in this figure and
Fig. \ref{irreducible-vertex.fig} represent equilibrium Green's functions 
(unperturbed by the external field $U$). The vertex $K$ for the approximations considered
here is displayed in Fig. \ref{irreducible-vertex.fig}.}
\label{susceptibility.fig}
\end{figure}

\begin{figure}[ht]
\centerline{\includegraphics[scale=0.5,angle=0,trim=00 20 00 00,clip=true]
{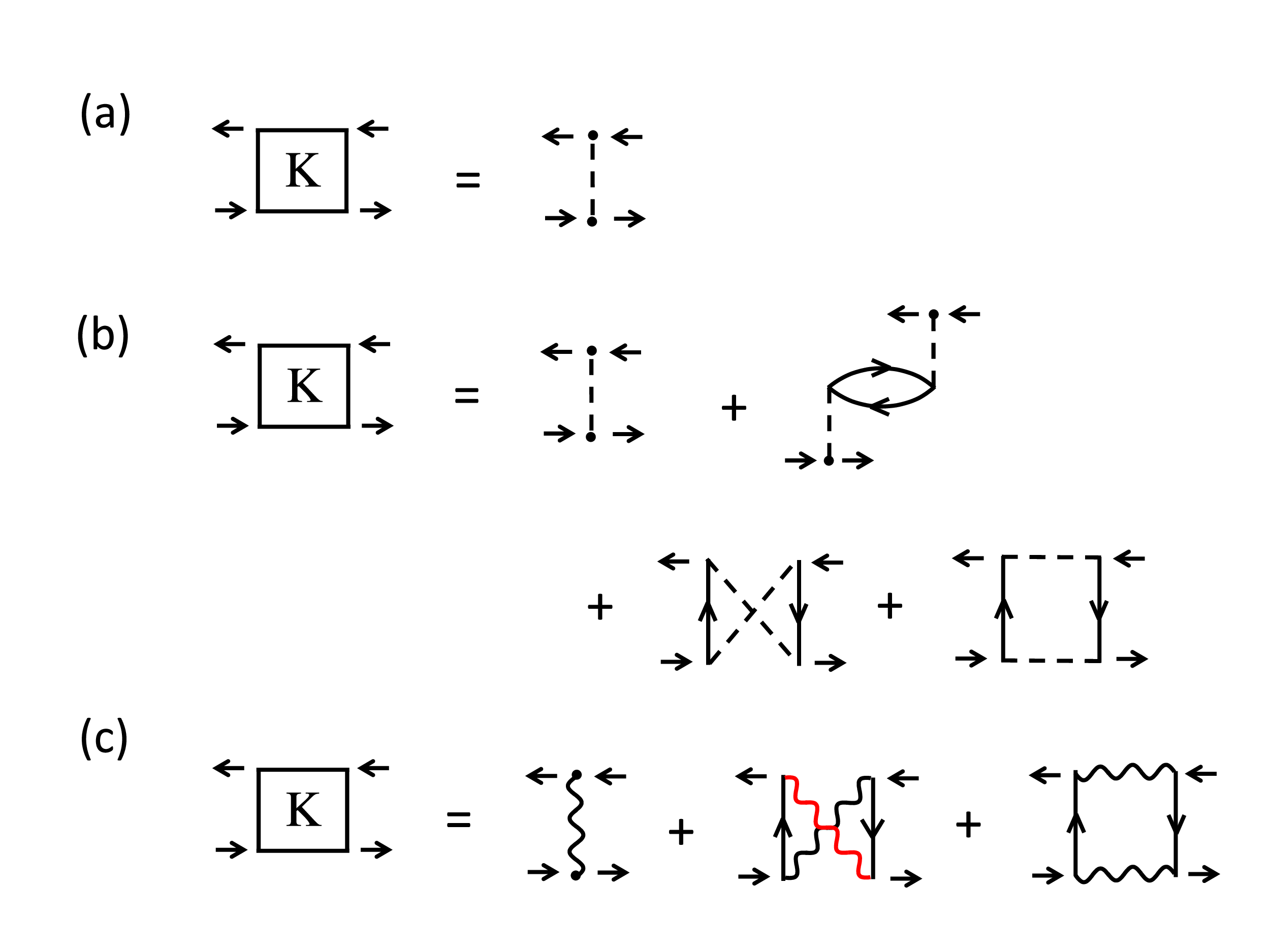}}
\caption{The diagrams representing the vertex $K$ in Fig. \ref{susceptibility.fig} that
are equivalent to the three self-energy approximations used in this paper: 
(a) Hartree-Fock, (b) second Born, and (c) ring diagrams. One interaction line in (c)
is drawn in red to help make the diagram clearer.
}
\label{irreducible-vertex.fig}
\end{figure}

In the conventional formulation of linear (density) response theory (e.g.
\cite{fetter-walecka.71}), a retarded density correlation function,
made up of equilibrium single-particle Green's functions, is calculated.
For our nuclear matter system driven by the perturbation 
$(1 / V ) U ( t )\sum\limits_{{\bf k} \sigma \tau} a^{\dag}_{{\bf k}+{\bf q} \sigma \tau}
a_{{\bf k} \sigma \tau}$($V$ being the normalization volume), the
first-order density response carrying momentum ${\bf q}$,
$\delta n({\bf q},t) = \sum\limits_{{\bf k} \sigma \tau} 
\langle a^{\dag}_{{\bf k}-{\bf q} \sigma \tau} ( t ) a_{{\bf k} 
\sigma \tau} ( t ) \rangle$, is written as  
\begin{equation}
\delta n({\bf q},t) = \frac {1} {\hbar} \int^{\infty}_{- \infty} 
\Pi^{R} ( {\bf q} , t - t') U ( t' )
\end{equation}
where the retarded density correlation $\Pi^{R} ( {\bf q} , t - t' )$ is defined as
\begin{equation}
\Pi^{R} ( {\bf q} , t - t' ) = 
- \frac {i} {V} \theta ( t - t' ) \sum\limits_{\substack{{\bf k} {\bf k'} \\ \sigma \tau}} 
\langle [ a^{\dag}_{{\bf k}-{\bf q} \sigma \tau} ( t ) a_{{\bf k} \sigma \tau} ( t), 
a^{\dag}_{{\bf k}'+{\bf q} \sigma \tau} ( t' ) a_{{\bf k}' \sigma \tau} ( t' ) ]
\rangle_{0}
\end{equation}
where the subscript $0$ indicates that the expectation value is taken in the
equilibrium (unperturbed by $U(t)$) state. To calculate $\Pi^{R} ( {\bf q} , t - t' )$,
it is related to the equilibrium Keldysh-time-ordered correlation function
\begin{equation}
\Pi ( {\bf q} , {\bar t} , {\bar t}' ) = - \frac {i} {V} \sum\limits_{\substack{{\bf k}
{\bf k}' \\ \sigma \tau}} \langle T_C [ a^{\dag}_{{\bf k}-{\bf q} \sigma \tau} 
( {\bar t} ) a_{{\bf k} \sigma \tau} ( {\bar t} ) a^{\dag}_{{\bf k}'+{\bf q} 
\sigma \tau} ( {\bar t}' ) a_{{\bf k}' \sigma \tau} ( {\bar t}' ) ] \rangle_{0}
\end{equation}
which may be evaluated in diagrammatic perturbation theory. Figs. 13 and 14 show 
the diagrams representing $\Pi ( {\bf q} , {\bar t} , {\bar t}' )$ that correspond
to the three approximations to the perturbed self energy used in this paper. Our
diagrammatic derivation of $\Pi({\bf q},{\bar t},{\bar t}')$ is similar to that in
\cite{bay62}. 
We note again that all the single-nucleon Green's 
functions in these two figures are evaluated in the equilibrium state. Neglecting
the quantity $L$ (represented by the rectangle labeled by $L$ in Fig.  
\ref{susceptibility.fig}) would yield the random phase approximation.  Containing 
the effects of exchange, correlations and collisions, $L$ is written as the 
solution of a Bethe-Salpeter equation (third equation in Fig.
\ref{susceptibility.fig}) driven by a vertex $K$. Each approximation for the 
perturbed self energy $\Sigma_{10}$ in our formulation corresponds to an equivalent 
selection of diagram set for $K$.  These diagrams are drawn in 
Fig.  \ref{irreducible-vertex.fig}. For each choice of diagrams for 
$K$, the single-nucleon Green's functions in the diagrams are dressed by the equilibrium 
self energy $\Sigma_{00}$ evaluated to the same level of approximation (Hartree-Fock, 
second Born, and ring diagrams) as that for the chosen $\Sigma_{10}$.  

\newpage
\section{Acknowledgements}

One of us (HSK) wishes to thank The University of Arizona and in
particular the Department of Physics for providing office space and
access to computer facilities.
	This work has not been supported by any external agency.

\end{document}